\documentclass[aip,10pt,floatfix,nofootinbib,twocolumn]{revtex4-1}
\usepackage{graphicx}
\usepackage{dcolumn}
\usepackage{bm}
\usepackage{amsmath}
\usepackage{amssymb}
\usepackage{psfrag}
\usepackage{color}
\usepackage{lipsum}
\newcommand{\parallelslam}{\mathbin{\!/\mkern-5mu/\!}}

\begin{document}

\title{
Cholesteric and screw-like nematic phases in 
systems of helical particles
}
\author{Giorgio Cinacchi}
\email{giorgio.cinacchi@uam.es}
\affiliation{
Departamento de F\'{i}sica Te\'{o}rica de la Materia Condensada, \\
Instituto de F\'{i}sica de la Materia Condensada (IFIMAC) and \\
Instituto de Ciencias de Materiales "Nicol\'{a}s Cabrera", \\ 
Universidad Aut\'{o}noma de Madrid,
Campus de Cantoblanco, 28049 Madrid, Spain
}

\author{Alberta Ferrarini}
\email{alberta.ferrarini@unipd.it}
\affiliation{Dipartimento di Scienze Chimiche, \\
Universit\`{a} di Padova, via F. Marzolo 1, 35131 Padova, Italy}

\author{Achille Giacometti}
\email{achille.giacometti@unive.it}
\affiliation{Dipartimento di Scienze Molecolari e Nanosistemi, \\
Universit\`{a} Ca' Foscari di Venezia,
via Torino 155, 30172 Venezia Mestre,
Italy}

\author{Hima Bindu Kolli}
\affiliation{
Department of Chemistry, \\ 
University of Oslo, 
Postboks 1033 Blindern, 0315 Oslo, Norway}

\date{\today}

\begin{abstract}
Recent numerical simulations of hard helical particle systems unveiled 
the existence of a novel chiral nematic phase, termed screw-like, 
characterised by the helical organization of the particle C$_2$ symmetry axes round 
the nematic director with periodicity equal to the particle pitch. 
This phase forms at high density and 
can follow a less dense uniform nematic phase, 
with relative occurrence of the two phases depending on the helix morphology. 
Since these numerical simulations were conducted under three-dimensional periodic boundary conditions,
two questions could remain open.
Firstly, the real nature of the lower density nematic phase, 
expected to be cholesteric.
Secondly,   
the influence that the latter, once allowed to form, may have on 
the existence and stability of the screw-like nematic phase.
To address these questions, we have performed 
Monte Carlo and molecular dynamics numerical simulations of 
helical particle systems confined between two parallel repulsive walls. 
We have found that removal of the periodicity constraint along one direction allows 
a relatively-long-pitch cholesteric phase to form, in lieu of the uniform nematic phase, with 
helical axis perpendicular to the walls  
while the existence and stability of the screw-like nematic phase 
are not appreciably affected by this change of boundary conditions.
\end{abstract}

\maketitle 

\section{Introduction}
\label{sec:introduction}

The propagation of chirality from the microscopic to the macroscopic scale is 
an issue of importance, both for fundamental science and potential applications. 
One very interesting case is represented by chiral particles experiencing only steric interactions.
Within this class, one of the most natural and simplest model is the hard helix.
More than 40 years ago, \cite{straley} 
hard helices of sufficiently high aspect ratio were predicted 
to form a cholesteric phase ($\mathsf{N}_c^*$),  a nematic liquid-crystal \cite{dg} phase
in which the main (usually long) particle axes are 
locally preferentially aligned parallel to one another and 
the average alignment axis, the nematic director ($\widehat{{n}}$),  revolves in 
a helical fashion round a perpendicular axis ($\widehat{\mathbf{h}}$) with 
a half-pitch (${\cal P}/2$) a few orders of magnitude larger than any particle dimension. 
Within the same framework, \cite{straley} 
a relationship was proposed between 
the handedness of the $\mathsf{N}_c^*$ phase and 
the morphology  (not simply the handedness) of the constituent helices. 
More recently, we addressed the phase behaviour of hard helical particles by 
Onsager-like (density functional) theory and Monte Carlo numerical simulation. 
\cite{Frezza13,Kolli14a,Frezza14,Kolli14b,Kolli16}
We found a rich liquid-crystal polymorphism in terms of the helix morphology.  
Particularly noteworthy was the observation of a novel chiral nematic phase,
termed screw-like  ($\mathsf{N}^*_s$), 
distinct from the $\mathsf{N}^*_c$ phase. 
In the $\mathsf{N}^*_s$ phase, $\widehat{{n}}$ $\parallel$ $\mathbf{\widehat{h}}$ and 
it is a transverse  director ($\widehat{\mathbf{m}}$) that 
revolves in a helical fashion round $\widehat{\mathbf{h}}$ with a pitch equal to that of the particle.
While the $\mathsf{N}^*_c$ phase can be exhibited by 
any non-racemic system of  chiral nematogenic particles, 
the $\mathsf{N}^*_s$ phase is special to helical particles and 
its formation sensitively depends on the helix morphology. 
It is this $\mathsf{N}^*_s$ phase that was observed 
in experiments on colloidal suspensions of helical flagella \cite{barry} and 
its possible existence in dense DNA solutions was also interestingly hypothesized \cite{manna}.

In our previous numerical simulations, \cite{Frezza13,Kolli14a,Frezza14,Kolli14b,Kolli16}  
three-dimensional periodic boundary conditions (3D-PBC) were used.
While entirely compatible with the $\mathsf{N}^*_s$  phase, 
3D-PBC can clearly be inadequate in the case of a $\mathsf{N}^*_c$ phase 
with a value of ${\cal P}/2$ orders of magnitude larger than 
computational box dimensions.\cite{memmerJCP} 
In our numerical simulations a uniform nematic ($\mathsf{N}$) phase was thus observed, 
which could precede the $\mathsf{N}^*_s$ phase on increasing density from the isotropic phase.  
Theoretical calculations made by us\cite{Frezza14,Kolli16},
reproduced by others \cite{Belli14,Dussi15,tortora}, 
confirm the expectation\cite{straley,Frezza13,Kolli14a,Kolli14b} that this $\mathsf{N}$ phase is actually cholesteric while
providing a prediction for the sign and the values of the corresponding $\cal{P}$.\cite{comment}
This could raise the question of the actual observability of 
a $\mathsf{N}_c^*$ phase for helical particles together with that of 
the actual existence and stability of the 
$\mathsf{N}^*_s$ phase with respect to a change in the boundary conditions that would
allow a proper  $\mathsf{N}^*_c$ phase to form.\cite{FaradayDisc2016}
The aim of the present study is to address both these points.

The numerical simulation of cholesteric liquid crystals poses specific problems. 
\cite{Frenkel2013}
In the last few years, 
this challenging objective has been particularly pursued. \cite{Melle13,Dussi16,wensinkmd} 
Hard helical particles were marginally addressed in Ref. \onlinecite{Dussi16} 
where difficulties arising in this case, related to the length of 
$\cal{P}$ and the presence of the $\mathsf{N}^*_s$ phase,  were remarked.

Here, we further clarify the nature of the nematic phases formed by helical particles,
removing any doubt about possible artefacts deriving by the use of 3D-PBC. 
By using two different independent numerical simulation methods, Monte Carlo and molecular dynamics, 
we will explicitly show the existence of a $\mathsf{N}_c^*$ phase for not-too-curly helical particles,
thus supporting the suggestion \cite{Frezza13,Kolli14a,Frezza14,Kolli14b,Kolli16}
that the previously observed $\mathsf{N}$ phase will actually turn cholesteric once 3D-PBC had been suitably removed
as well as we confirm the existence and stability of the $\mathsf{N}_s^*$ phase 
for sufficiently curly helical particles also under
these other boundary conditions. \cite{noteonyork} 

In the next Section \ref{sec:model} we will recall the model and 
describe the numerical simulation protocols.
Section \ref{sec:results}
presents the results of this study 
preceded by a summary of past results to cast it in the appropriate perspective.
Section \ref{sec:conclusions} summarises the main findings.

\section{Model and computational details}
\label{sec:model}

\begin{figure}
\centering
\psfrag{rog}{$\mathbf{r}_{\circ}$}
\psfrag{uu}{$\widehat{\mathbf{u}}$}
\psfrag{vv}{$\widehat{\mathbf{v}}$}
\psfrag{ww}{$\widehat{\mathbf{w}}$}
\psfrag{rr}{$r$}
\psfrag{pp}{$p$}
\psfrag{nn}{$ $}
\psfrag{mm}{$ $}
\includegraphics[scale=0.85]{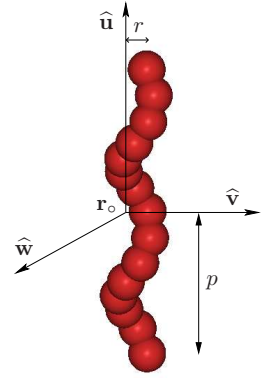}
\caption{
Illustration of a 
particle composed of
15 partially overlapping spherical beads of diameter $D$ 
whose centres are equidistantly  and rigidly arranged along 
a right-handed helical cord of length $L$, radius $r$ and pitch $p$.
The mechanical state of any such helical particle can be defined by 
specifying its reference frame with respect to the laboratory reference frame:
the position of the origin
$\mathbf{r}_{\circ}$ and 
the orientation of the three mutually perpendicular unit vectors:
$\mathbf{\widehat{u}}$, $\mathbf{\widehat{v}}$ and $\mathbf{\widehat{w}}$.
}
\label{singolelica}
\end{figure}

In keeping with previous works,\cite{Frezza13,Kolli14a,Kolli14b,Frezza14,Kolli16,dinamica} 
the model particles considered are 
made of 15 partially overlapping hard or soft-repulsive spheres of diameter $D$,
the unit of length, equidistantly and
rigidly arranged along a right-handed helix  
of fixed contour length $L$=10$D$, 
varying radius $r$ and pitch $p$ and whose  
long and short axes are, respectively, denoted as  
$\widehat{\mathbf{u}}$ and $\widehat{\mathbf{w}}$ (Fig. \ref{singolelica}).

In case the spherical beads are soft-repulsive,
two of them belonging to different  
particles interact via the repulsive part $V(r_{kl})$ of 
the common separation of the Lennard-Jones potential energy function:\cite{WCA}
\begin{eqnarray}
V(r_{kl})=\left\{ 
\begin{array}{lr}
4\epsilon \left[\left(\frac{D}{r_{kl}}\right)^{12}-\left(\frac{D}{r_{kl}}\right)^6 + \frac{1}{4}\right] 
& : r_{kl} \le \sqrt[6]{2}D \\
0 & : r_{kl} > \sqrt[6]{2}D
\end{array}
\right.
\end{eqnarray}
with $r_{kl}$ the distance between the centres of beads $k$ and $l$ and
$\epsilon$ the unit of energy.

For an on-lattice model, 
self-determined spiralling boundary conditions were devised as
the solution to 
the problem of dealing with phases whose structure is characterised by 
a length scale that may be incommensurate to 
the dimensions of the lattice.\cite{spiralling,cholestlatt}
In the same remarkable work,\cite{spiralling} 
it was stated that the next best solution to this problem would be to employ free boundary conditions.
For an off-lattice model, self-determined spiralling boundary conditions cannot be properly employed. 
Then, we have followed the
suggestion \cite{Melle13} that a $\mathsf{N}^*_c$ phase can be investigated in a numerical simulation by confining
the system between two suitable $\widehat{\mathbf{h}}$-perpendicular walls. 
The price to pay is to deal with a confined rather than bulk system.
However, if the slab is sufficiently thick, 
the properties of the confined system sufficiently far away from the walls are
equivalent to the corresponding properties of the bulk system.

Thus, the adopted computational protocol consists of two parts. 
Firstly, it is observed  that 
confining 
between two parallel hard or soft-repulsive walls  
a sufficiently large and potentially cholesterogenic system,
previously studied under 
3D-PBC and
seen to be nematogenic, does not alter
its properties away from the walls:
it only makes  $\widehat{n}$ twist and a $\mathsf{N}^*_c$ phase result. 
Secondly, it is observed that 
the $\mathsf{N}^*_s$ phase 
either remains stable even under confinement,
when a $\mathsf{N}^*_c$ phase could form, 
or
spontaneously develops when starting from a cholesteric configuration.

(Enantiomerically) pure systems of N$\in[1500,5000]$  
(freely translating and rotating) hard or soft-repulsive helical particles, 
identified by the pair ($r$,$p$), 
were placed between 
two fixed parallel, respectively hard or soft-repulsive, flat walls
perpendicular to the $x$ axis of
the laboratory reference frame.
In case of soft repulsive walls,
their interactions with a  
spherical bead $k$ was described via the potential energy function $W_{9/3}(r_k)$:
\begin{eqnarray}
W_{9/3}(r_{k})=\left\{ 
\begin{array}{lr}
\epsilon \left[\frac{2}{15} \left(\frac{D}{r_{k}}\right)^{9}-\left(\frac{D}{r_{k}}\right)^3 + 1\right] 
& : r_{k} \le \sqrt[6]{0.4}D \\
0 & : r_{k} > \sqrt[6]{0.4}D
\end{array}
\right.
\end{eqnarray}
with $r_k$ the distance of the spherical bead $k$ from a wall.
Such helical-particle--wall interactions promote planar alignment of the helical particles
close to the walls so that $\widehat{\mathbf{h}} \parallel x$.
 
The systems were investigated via, respectively, the Monte Carlo (MC) method\cite{MCorig}
or the molecular dynamics (MD) method.\cite{MDorig}
Both were run in the isobaric-isothermal (NPT) ensemble\cite{MCorignpt,MDorignpt,allenfrenkel}
for several values of pressure P measured:
in units $k_B {\rm T}/D^3$,
with $k_B$ the Boltzmann constant and T the absolute temperature,
in the MC case; in units $\epsilon/D^3$ with $k_B {\rm T}/\epsilon = 1$,
in the MD case.
Both the MC-NPT and MD-NPT calculations were
carried out 
under rectangular periodic boundary conditions 
along the $y$ and $z$ axes of the laboratory reference frame.\cite{Melle13}
The MC-NPT calculations were organised in cycles,
each consisting of
2N attempts to translate or rotate a randomly selected helical particle plus
an attempt to change the cuboidal computational box shape and volume by
independently changing the length of its $L_y$ or $L_z$ edge.\cite{Melle13}
The MD-NPT calculations were 
carried out with the program LAMMPS,\cite{LAMMPS} integrating
the equations of motion via a rigid-particle algorithm\cite{rigidalgo}
with a time-step $\tau=0.01\sqrt{mD^2/\epsilon}$, $m$
being the mass of a helical particle, 
with a semi-isotropic 
barostat, allowing the cuboidal computational box 
to fluctuate in the $yz$ plane, and  a thermostat whose 
damping  parameters to control pressure and temperature,
$\tau_{\rm P}$ and $\tau_{\rm T}$, were both equal to $\sqrt{mD^2/\epsilon}$.
Both MC-NPT and MD-NPT calculations  
were started: i) either from
a configuration equilibrated during previous numerical simulations and 
inserted between the two parallel hard or soft-repulsive walls 
taking care to remove those particles that happened to overlap with them; ii) or
a moderately dense orthorhombic lattice configuration with 
all helical-particle $\widehat{\mathbf{u}}$ axes perfectly aligned along the
$z$ axis; iii) or a configuration obtained as output of a run at a nearby value of pressure P.
Rather lengthy simulations had  to be employed. 
In the MC case, most runs were of 15 or more million MC cycles.
In the MD case, the trajectories were up to 320 million time-step long.

In the (production) runs 
averages of several quantities were accumulated and 
configurations stored for the subsequent analysis.
The calculated quantities include
the number density, $\varrho$,
the nematic order parameter,\cite{vieilliard} $S_2$,
as well as 
the angle, $\theta$, that 
the local nematic director, $\widehat{{n}}(x)$, forms with the $y$ axis,
as a function of $x$,
along with a few suitable pair correlation functions
to further ascertain the modulated-nematic nature of a phase.
From a linear fit of $\theta(x)$, whose slope coincides with $q$=$2\pi/{\cal P}$,
an estimate of $\cal P$ was obtained.
One pair correlation function is 
\begin{equation}
g_2^{\widehat {\mathbf{u}}} (x) = 
\left \langle \frac{ \sum_{i=1}^{\rm N} \sum_{j>i}^{\rm N}P_2\left( {\widehat{\mathbf{u}}}_i \cdot 
{\widehat{\mathbf{u}}}_j \right) \delta(x-x_{ij}) } { \sum_{i=1}^{\rm N} \sum_{j>i}^{\rm N} \delta(x-x_{ij})} \right \rangle,
\end{equation}
with $P_n()$ the $n$th-order Legendre polynomial, 
$\delta()$ the standard $\delta$-function, $x_{ij}$ the distance between particle $i$ and $j$,
$\mathbf{r}_{ij}$, 
resolved along the $x$ axis and  $\left \langle \right \rangle$ representing an average 
over configurations. This function 
is expected 
to behave as  $S_2^2 P_2(\cos\left(qx\right))$ if a cholesteric ordering is present
and be essentially flat otherwise.
One more correlation function is $\displaystyle g_1^{\mathbf{w}}\left(r_{\parallel},x\right)$,
the suitable generalisation of the pair correlation function 
$\displaystyle g_1^{\widehat {\mathbf{w}}} \left( r_{\parallel} \right) $ previously defined.\cite{Kolli14a,Kolli14b}
Rather than explicitly considering this two-variable function,
it proves convenient to consider either the function 
$\displaystyle G_1^{\widehat {\mathbf{w}}}(x)=g_1^{\mathbf{w}}\left(0,x\right) 
\equiv \max\limits_{r_{\parallel}}g_1^{\mathbf{w}}\left(r_{\parallel},x\right)$ or the function
$g_{1,\Delta}^{\widehat {\mathbf{w}},\widehat{\mathbf{u}}} \left( r_{\parallelslam} \right)$
defined as:
\begin{widetext}
\begin{equation}
g_{1,\Delta}^{\widehat {\mathbf{w}},\widehat{\mathbf{u}}} \left( r_{\parallelslam} \right) = 
\left \langle \frac{ \sum_{i=1}^{\rm N} \sum_{j>i}^{\rm N}P_1\left( {\widehat{\mathbf{w}}}_i \cdot 
{\widehat{\mathbf{w}}}_j \right) \left[ (x_{ij}-\Delta)<0 \right] \delta \left(r_{\parallelslam}-\mathbf{r}_{ij} \cdot
\widehat{\mathbf{u}}_i \right) } { \sum_{i=1}^{\rm N} \sum_{j>i}^{\rm N}
\left [ (x_{ij}-\Delta)<0\right]
\delta\left(r_{\parallelslam}-\mathbf{r}_{ij} \cdot
\widehat{\mathbf{u}}_i \right)} \right \rangle,
\end{equation}
\end{widetext}
with $\left[(x_{ij}-\Delta)<0\right]$ an Iverson bracket. 
This variant of $\displaystyle g_1^{\widehat {\mathbf{w}}} \left( r_{\parallel} \right)$ 
is expected
to develop a well-defined $p$-periodic cosinusoidal form if
a screw-like ordering is overall present and be essentially flat otherwise. 
In case $g_{1,\Delta}^{\widehat {\mathbf{w}},\widehat{\mathbf{u}}} \left( r_{\parallelslam} \right)$ reveals a presence of screw-like ordering, the function 
$\displaystyle G_1^{\widehat {\mathbf{w}}}(x)$ can further specify  
its dependence on $x$.
Together, the pair correlation function $g_2^{\widehat {\mathbf{u}}} (x)$ and
the duet
$\displaystyle G_1^{\widehat {\mathbf{w}}}(x)$ 
-- $g_{1,\Delta}^{\widehat {\mathbf{w}},\widehat{\mathbf{u}}} \left( r_{\parallelslam} \right)$  
reveal whether the two
types of modulated-nematic ordering are present, either jointly or separately.

 \section{Results}
  \label{sec:results}
\subsection{Systems under investigation}
\label{subsec:summary}
\begin{figure*}
\centering
\psfrag{PD3kT}{P$D^3 / k_b$T}
\psfrag{eta}{$\eta$}
\psfrag{iso}{$\mathsf{I}$}
\psfrag{nem}{$\mathsf{N}$}
\psfrag{nems}{$\mathsf{N}_s^*$}
\psfrag{SmA}{$\mathsf{S}_{A,s}^*$}
\psfrag{SmB}{$\mathsf{S}_{B,s}^*$}
\psfrag{SmBp}{$\mathsf{S}_{B,p}$}
\psfrag{r=0.2D,p=8D}{$r=0.2D$,$p=8D$}
\psfrag{r=0.2D,p=9.92D}{$r=0.2D$,$p=9.92D$}
\psfrag{r=0.2D,p=4D}{$r=0.2D$,$p=4D$}
\psfrag{r=0.4D,p=4D}{$r=0.4D$,$p=4D$}
\includegraphics[scale=0.85]{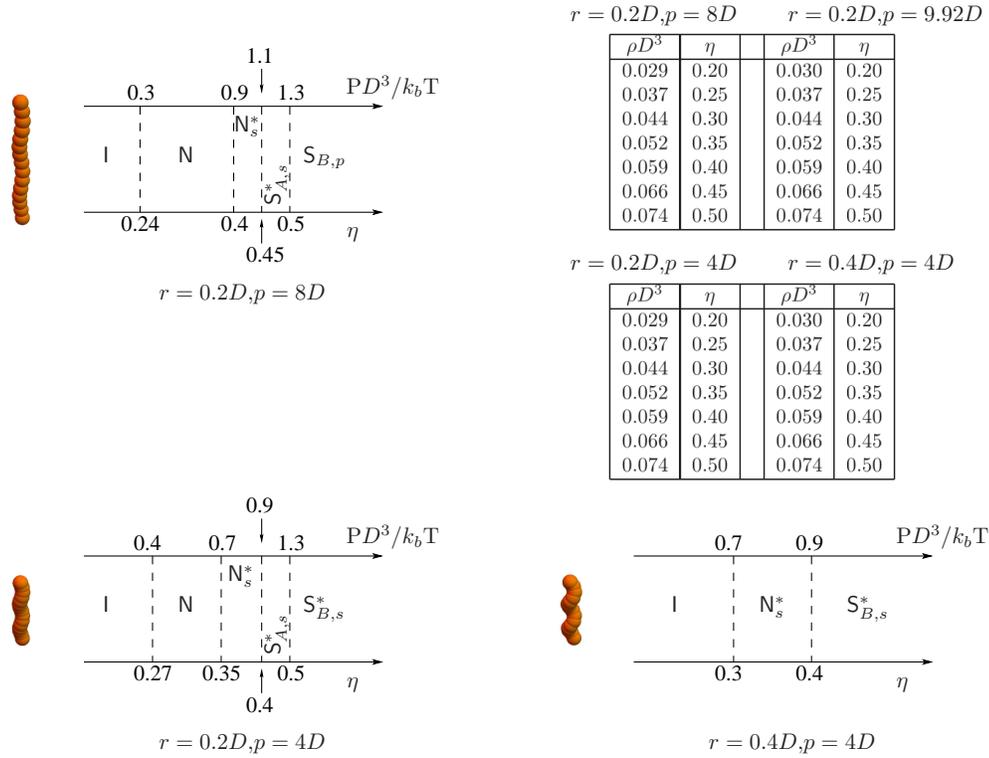}
  \caption{Summary of the phase diagrams for three different helix morphologies, 
as a function of the volume fraction $\eta=\varrho v$, with $v$ the volume of a helical particle
whose image is included: 
(top left) $r=0.2D$ and $p=8D$; (bottom left) $r=0.2D$ and $p=4D$;    
(bottom right) $r=0.4D$ and $p=4D$. 
Here, $\mathsf{I}=$isotropic phase, 
$\mathsf{N}=$ uniform nematic, 
$\mathsf{N}^*_s=$screw-like nematic, 
$\mathsf{S}^*_{A,s}=$ screw-like smectic A, 
$\mathsf{S}^*_{B,s}=$ screw-like smectic B. 
$\mathsf{S}_{B,p}=$ polar smectic B. 
The relations between 
$\eta$ and $\rho D^3$ for these three cases as well
as the case with $r=0.2D$ and $p=9.92D$ are explicitly given (top right).}
    \label{fig:summary}
    \end{figure*}
The present work builds upon past extensive numerical simulations, employing 3D-PBC, of systems of
hard helical particles with various parameters ($r, p$).
\cite{Frezza13,Kolli14a,Kolli14b,Kolli16} 
Here, we have focused on three representative cases 
which were found to exhibit different propensity to form $\mathsf{N}$ and $\mathsf{N}_s^*$ phases,
as schematically summarised in Figure \ref{fig:summary}. 

For helical particles with small $r$ and relatively long $p$
($r=0.2D$ and $p=8D$),  a broad-range $\mathsf{N}$ phase is found, 
followed, at progressively higher density, by a $\mathsf{N}^*_s$ phase, on a very narrow region,
and    
screw-like smectic A and B phases.
For helical particles with the same $r$ and a smaller $p$ ($r=0.2D$ and $p=4D$), 
one finds both $\mathsf{N}$ and $\mathsf{N}^*_s$ phases to be present in respective regions
of comparable width. 
For helical particles with a larger $r$ and the same $p$
($r=0.4D$ and $p=4D$) the $\mathsf{N}$ phase is absent and  
a region of $\mathsf{N}^*_s$ phase is found.

These findings were rationalized \cite{Kolli14a,Kolli14b} in terms of 
an entropy gain triggered by the coupling between 
translation of the particles along  $\widehat{n}$ and
rotation round  their own $\widehat{\mathbf{u}}$. 
On increasing their curliness, 
neighbouring parallel helical particles tend to 
mutually interlock their grooves, 
thus restraining rotation round their  $\widehat{\mathbf{u}}$. 
This rotational entropy loss can be compensated by 
a translational motion along $\widehat{n}$ in a way similar to a screw: 
hence the name given to the new chiral nematic phase.

  \subsection{$r$=0.2$D$ and $p$=9.92$D$}
  \label{subsec:r02p992}
\begin{figure}
\centering
\psfrag{densityreducedunits}{$\varrho D^{3}$}
\psfrag{pressurereducedunits}{$\frac{{\rm P}D^{3}}{k_B {\rm T}}$}
\psfrag{nematicorderparameter}{$S_2$}
\psfrag{isotropic}{\tiny{isotropic}}
\psfrag{nematicpbc} {\tiny{nematic (pbc)}}
\psfrag{cholesterichw} {\tiny{cholesteric (hw or sw)}}
\psfrag{pbc}{\tiny{pbc}}
\psfrag{hw}{\tiny{hw}}
\psfrag{sw}{\tiny{sw}}
\includegraphics[scale=0.75]{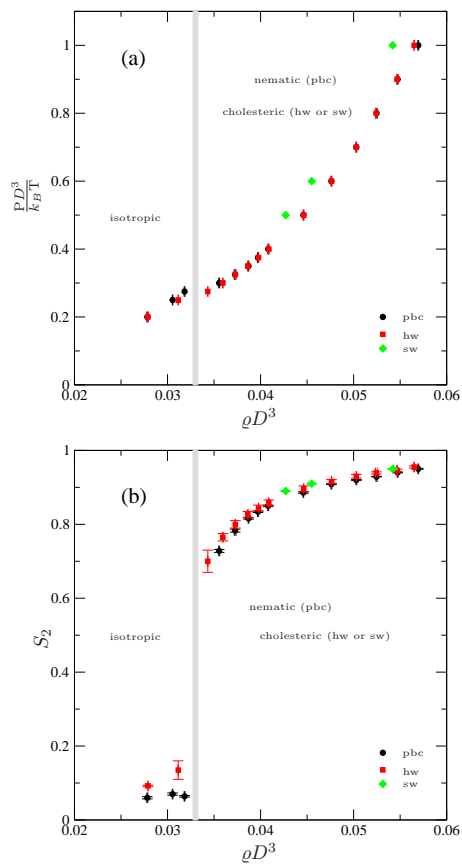}
\caption{Behaviour of dimensionless pressure 
$\displaystyle \frac{{\rm P}D^{3}}{k_B {\rm T}}$ (a) and nematic
order parameter $S_2$ (b) as a function
of dimensionless number density $\varrho D^{3}$
for a system of hard helical particles with $r=0.2 D$ and $p=9.92 D$
as obtained in MC numerical simulations
employing  3D-PBC
(pbc, black circles) or confining the system between two parallel hard walls (hw,
red squares). 
While in the former case a uniform nematic phase is formed, 
in the latter a cholesteric phase is observed. For comparison,
a few data obtained for a system of soft-repulsive
helical particles confined between two parallel soft-repulsive walls (sw, green
diamonds) which forms a cholesteric phase, have been also included.
In both panels, the gray slender box delimits either of these nematic
phases and the coexistent isotropic phase.
}
\label{figuraeospius2}
\end{figure}

\begin{figure}
\centering
\psfrag{thdx/rad} {$\theta/$rad}
\psfrag{x/D} {$x/D$}
\includegraphics[scale=0.75]{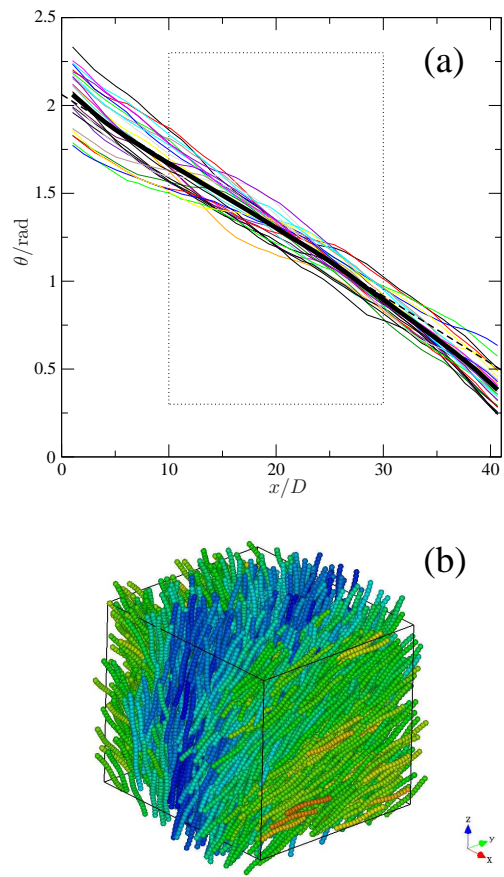}
\caption{
(a) Behaviour of 
the angle $\theta$ that the local nematic director forms 
with the $y$ axis  as a function of $x$ 
for a system of hard helical particles with $r=0.2 D$ and $p=9.92 D$ 
at  P$D^3/k_B$T=0.5. 
The various coloured thin full curves shown are each an average
over 200, one-thousand-MC-cycle separated, 
configurations while the black thick full curve
is their global average and the black thick dashed line is the
linear fit to the latter taking into account only the region
enclosed by the dotted rectangle.
(b) Image\cite{QMGA}
of the cholesteric phase 
for a system of hard helical particles with $r=0.2 D$ and $p=9.92 D$
 at  P$D^3/k_B$T=0.5.
The hard helical particles are coloured according to the angle 
that their $\widehat{\mathbf{u}}$ axes forms with 
the eigenvector corresponding to the largest eigenvalue
of the global nematic order matrix.\cite{vieilliard}
}
\label{figth}
\end{figure}

The examination of the obtained results starts 
with those for systems of helical particles having $r$=0.2$D$ and $p$=9.92$D$. 
This value of $p$ is such that, for
a contour length $L$, it corresponds to 
a single helical turn. 
These helical particles, nearly straight rods, 
have a wider $\mathsf{N}$ phase as 
compared to the $p=8D$ counterpart (Fig. \ref{fig:summary}) and
are expected to form the $\mathsf{N}_c^*$ only. \cite{Frezza13,Kolli14a,Kolli14b,Kolli16,Anda}
In order to confirm this, Fig. \ref{figuraeospius2} reports the equation of state (EoS) and 
the density dependence of $S_2$
as obtained for a bulk system of hard helical particles 
employing 3D-PBC
and for a system of the same particles 
confined between two parallel hard walls. 
Included are also a few data obtained for a system of soft-repulsive
helical particles confined between two parallel soft-repulsive walls. 
In all cases, the average values of 
$S_2$ correspond to the average values
of this quantity
calculated by  discarding the contribution of particles 
closer than 10$D$ to either wall.
The choice of this distance has been made by plotting the number density $\varrho(x)$ and   
nematic order parameter $S_2(x)$ profiles 
with respect to $x$, and by 
observing that, farther than that distance,
confinement effects had reasonably faded away.
From Fig. \ref{figuraeospius2}, it can be seen that: i) the isotropic-nematic (cholesteric) phase transition is slightly shifted to a lower
value of density as compared to the case $r=0.2D$, $p=8D$, as expected in view of 
the increased aspect ratio of the helical particles; ii)
both the EoS and $S_2$ versus $\varrho$ graphs for the confined system match
the corresponding graphs for the bulk system.
However, the latter is in the $\mathsf{N}$ phase whereas the former relatively
promptly forms a $\mathsf{N}^*_c$ phase. 
Fig. \ref{figth} shows the behaviour of $\theta(x)$, 
calculated at P$D^3/k_B$T=0.5, along with an image of
a configuration in the $\mathsf{N}_c^*$ phase at the same value of dimensionless pressure.
\begin{figure*}
\centering
\psfrag{rD3}{$\varrho D^{3}$}
\psfrag{q 2pisuP}{$qD=\frac{2\pi}{{\cal P}}D$}
\psfrag{nematicorderparameter}{$S_2$}
\psfrag{isotropic}{\tiny{isotropic}}
\psfrag{nematicpbc} {\tiny{nematic (pbc)}}
\psfrag{ cholesteric } {\tiny{cholesteric }}
\psfrag{xDrpD}{$x/D; r_{\parallelslam}/D$}
\psfrag{g2rpg1srp}{$g_2^{\widehat {\mathbf{u}}} (x); g_{1,\Delta}^{\widehat {\mathbf{w}},\widehat{\mathbf{u}}} \left( r_{\parallelslam} \right)$}
\includegraphics[scale=0.7]{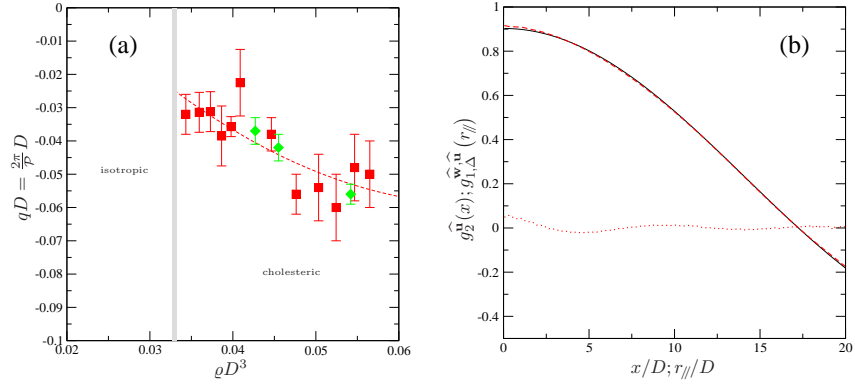}
\caption{(a) Behaviour of dimensionless wavevector
$\displaystyle q D=\frac{2\pi}{{\cal P}}D$ 
as a function of 
dimensionless number density $\varrho D^{3}$ (red squares)
for a system of hard helical particles with $r=0.2D$ and $p=9.92D$.
The red dashed line is a quadratic fit to the data,
meant as a mere guide to the eye. 
Included are also a few data obtained for a system
of soft-repulsive helical particles confined between
two parallel soft-repulsive walls (green diamonds).
The gray slender box delimits 
the cholesteric phase and the coexistent isotropic phase. 
(b) pair correlation functions $g_2^{\widehat {\mathbf{u}}} (x)$ (red dashed line) and 
$g_{1,\Delta}^{\widehat {\mathbf{w}},\widehat{\mathbf{u}}} \left( r_{\parallelslam} \right)$ (red dotted line) 
for a system of hard helical particles with $r=0.2D$ and $p=9.92D$ at P$D^3/k_B$T=1.
The black thinner full line is a fit of $g_2^{\widehat {\mathbf{u}}} (x)$ with
the function $S_2^2 P_2\left(\cos \left(qx\right) \right)$ with 
$S_2=0.950$ and $|q|D=0.0554$ as fitting parameters,
consistent with, respectively, 
the value of $S_2$ in Fig. \ref{figuraeospius2}(b)
and the absolute value of $qD$ in (a).
}
\label{figr02n1compost}
\end{figure*}
Discarding the contribution of 
the slices closer than 10$D$ to either wall,
linear fits to the various calculated $\theta(x)$'s were performed.
The values
of $q$ thus estimated
are given in Fig. \ref{figr02n1compost}(a)
as a function of $\varrho$.
The negative sign of $q$ indicates a left-handed $\mathsf{N}^*_c$ phase, as predicted by Straley \cite{straley} and Onsager-like  theory calculations \cite{Frezza14,Kolli16,Belli14,Dussi15,tortora} for right-handed hard helical particles with geometric parameters analogous to those of the system under investigation. The large error bars reflect wide fluctuations of the local director which do not allow a precise determination of $\cal{P}$.\cite{notapitch}
None the less, the  magnitude of $q$, corresponding to $\cal P$ in the range of 100-200 $D$, 
which decreases with increasing density, compares well to the theoretical calculations for hard helical particles with $r=0.2D$ and $p=8D$ in Ref. \onlinecite{Kolli16}.
The functions 
$g_2^{\widehat {\mathbf{u}}} (x)$ and 
$g_{1,\Delta}^{\widehat {\mathbf{w}},\widehat{\mathbf{u}}} \left( r_{\parallelslam} \right)$ at 
P$D^3/k_B$T $=1$ are shown in Fig. \ref{figr02n1compost}(b). 
Their behaviour confirms the cholesteric character of the nematic phase and
the expected absence of screw-like ordering  
[yet, an extremely tenuous short-range undulation is already visible in
the function $g_{1,\Delta}^{\widehat {\mathbf{w}},\widehat{\mathbf{u}}} \left( r_{\parallelslam} \right)$]:
these helical particles  are not sufficiently curly to promote, at these values of $\varrho$, the setting in of the latter type of ordering.
\subsection{$r=0.4D$ and $p=4D$}
\label{subsec:r04p4}
\begin{figure*}
\centering
\psfrag{x/D}{$x/D$}
\psfrag{T/rad}{$\theta$/rad}
\psfrag{rp /D}{$r_{\parallelslam}/D$}
\psfrag{g1p rp }{$g_{1,\Delta}^{{\widehat{\mathbf{w}}},{\widehat{\mathbf{u}}}}(r_{\parallelslam})$}
\includegraphics[scale=0.75]{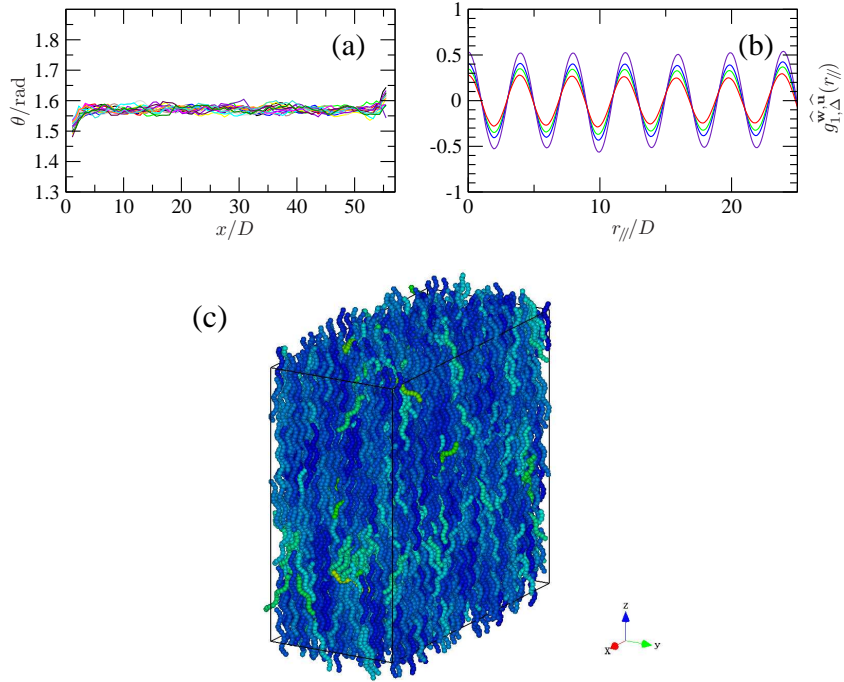}
\caption{For a system of hard helical particles with $r=0.4D$ and $p=4D$: 
(a) the angle $\theta$ that the local nematic director forms with 
the $y$ axis as a function of $x$ at P$D^3/k_B$T$=0.75$,
the various coloured thin full curves shown being each an 
average over 1000, one-thousand-MC-cycle separated, configurations;
(b) the pair correlation function 
$g_{1,\Delta}^{{\widehat{\mathbf{w}}},{\widehat{\mathbf{u}}}}(r_{\parallelslam})$
at  P$D^3/k_B$T$=0.7, 0.75, 0.8, 0.9$; 
(c) image\cite{QMGA} of the screw-like nematic phase 
at P$D^3/k_B$T$=0.75$. The hard helical particles are coloured according to the angle
that their $\widehat{\mathbf{u}}$ axes forms with
the eigenvector corresponding to the largest eigenvalue
of the global nematic order matrix.\cite{vieilliard}
}
\label{figr04p4composta}
\end{figure*}
The case of curly helical particles with $r=0.4D$ and $p=4D$ stands at the other extreme. 
Their nematic phase is screw-like, 
existing 
 in the interval P$D^3/k_B$T$\in (0.7,0.9)$, bounded by a less dense isotropic phase
and a denser screw-like smectic B phase. \cite{Kolli14a,Kolli14b,Kolli16}
This phase sequence, 
previously obtained employing 3D-PBC
for systems of hard \cite{Kolli14a,Kolli14b,Kolli16} and      
soft-repulsive\cite{dinamica} helical particles
with these values of $r$ and $p$, 
is confirmed  by the present MC-NPT calculations carried out under
confinement of two parallel hard walls.
Starting from a previously obtained configuration
carefully inserted in between the two parallel hard walls,
the $\mathsf{N}^*_s$ phase remains stable at P$D^3/k_B$T$=0.7,
0.8, 0.9$ in the course of runs up to 7 million-MC cycle--long (Fig.
\ref{figr04p4composta}),
while at P$D^3/k_B$T$ =0.6$ it transits 
directly to the isotropic phase.
\subsection{$r=0.2D$ and $p=4D$}
\label{subsec:r02p4}
\begin{figure}
\centering
\psfrag{xDrpD}{$x/D; r_{\parallelslam}/D$}
\psfrag{g2rpg1srp}{$g_2^{\widehat {\mathbf{u}}} (x); g_{1,\Delta}^{\widehat {\mathbf{w}},\widehat{\mathbf{u}}} \left( r_{\parallelslam} \right)$}
\includegraphics[scale=0.7]{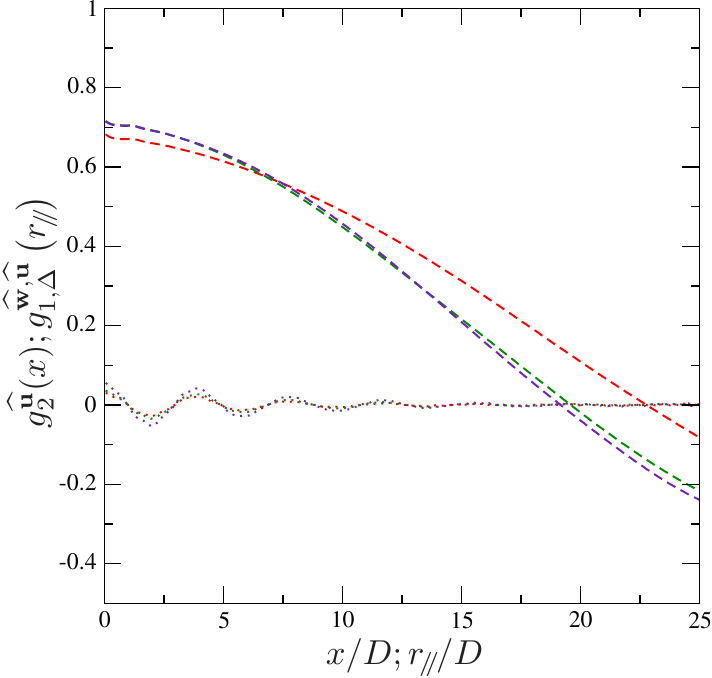}
\caption{Pair correlation functions $g_2^{\widehat {\mathbf{u}}} (x)$ (dashed line) and 
$g_{1,\Delta}^{\widehat {\mathbf{w}},\widehat{\mathbf{u}}} \left( r_{\parallelslam} \right)$ (dotted line) 
for a system of hard helical particles with $r=0.2D$ and $p=4D$ at
P$D^3/k_B$T=0.55 from an initial nematic configuration (red); 
0.6 from an initial nematic configuration (green); 
0.6 from an initial configuration taken from the run
at P$D^3/k_B$T=0.9, in its turn started from an orthorhombic lattice configuration (indigo).
}
\label{figP05506p}
\end{figure}
\begin{figure}[h!]
\centering
\psfrag{rho}{$\varrho D^3$}
\psfrag{S2}{$S_2$}
\psfrag{MC cycles}{MC cycles}
\psfrag{T/rad}{$\theta$/rad}
\psfrag{x/D}{$x/D$}
\psfrag{g1p rp }{$g_{1,\Delta}^{{\widehat{\mathbf{w}}},{\widehat{\mathbf{u}}}}(r_{\parallelslam})$}
\psfrag{rp/D}{$r_{\parallelslam}/D$}
\includegraphics[scale=0.7]{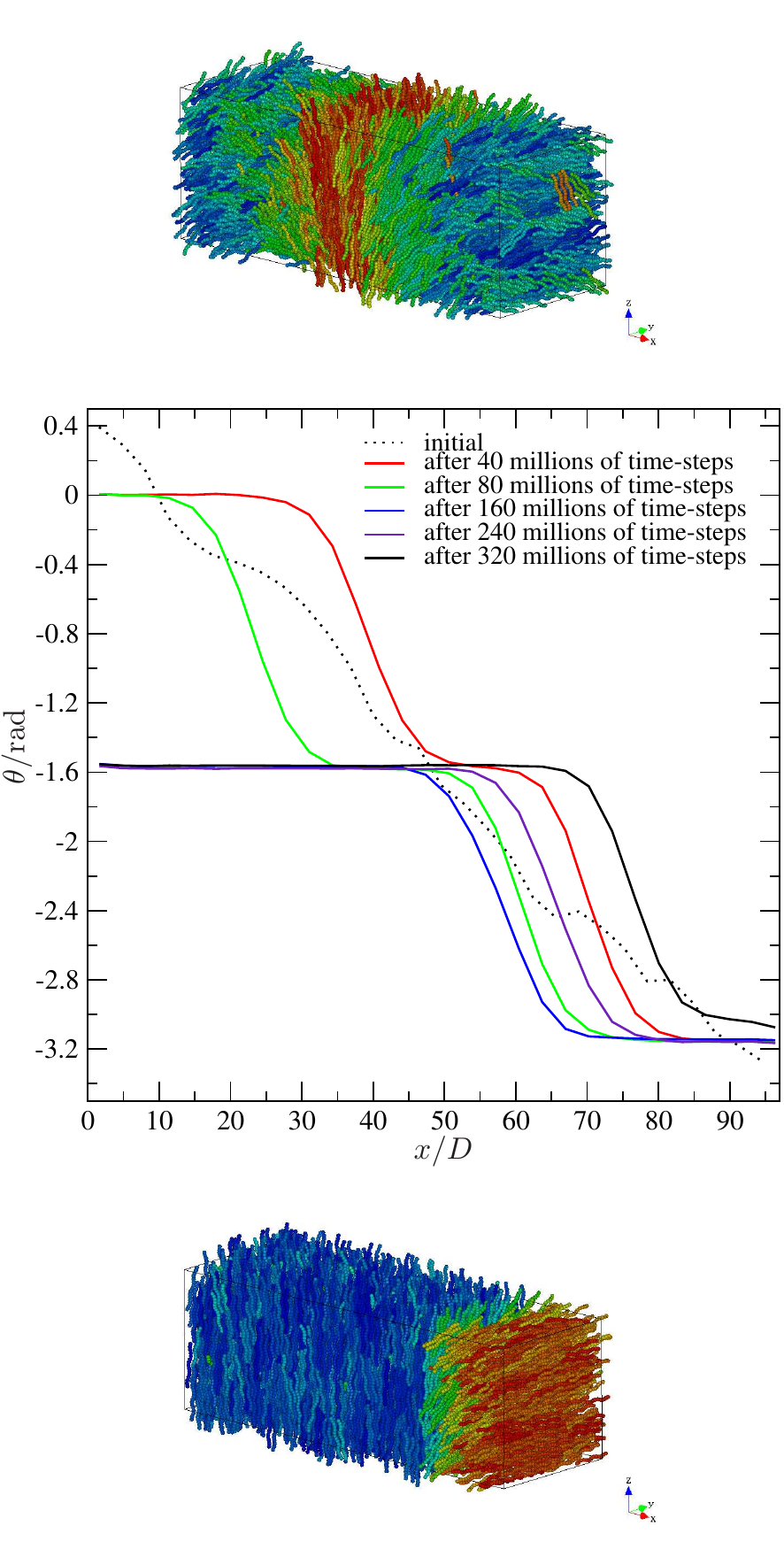}
\caption{Evolution of the local nematic director $\theta$ 
as a function of $x$ at several moments of the MD-NPT calculation
carried out on a system of soft-repulsive helical particles
with $r=0.2D$ and $p=4D$; apart from the initial,
each profile is an average over 10 million of time-steps
the final instant contributing to this average being that one indicated in the label.
Images\cite{QMGA} are shown of the initial (top) and final (bottom)
configuration. 
The soft-repulsive helical particles are coloured according to the angle
that their $\widehat{\mathbf{u}}$ axes forms with
the eigenvector corresponding to the largest eigenvalue
of the global nematic order matrix.\cite{vieilliard}
}
\label{r02p4emdcompost}
\end{figure}
\begin{figure*}[h!]
\centering
\psfrag{rho}{$\varrho D^3$}
\psfrag{S2}{$S_2$}
\psfrag{MC cycles}{MC cycles}
\psfrag{T/rad}{$\theta$/rad}
\psfrag{x/D}{$x/D$}
\psfrag{g1p rp }{$g_{1,\Delta}^{{\widehat{\mathbf{w}}},{\widehat{\mathbf{u}}}}(r_{\parallelslam})$}
\psfrag{rp/D}{$r_{\parallelslam}/D$}
\includegraphics[scale=0.75]{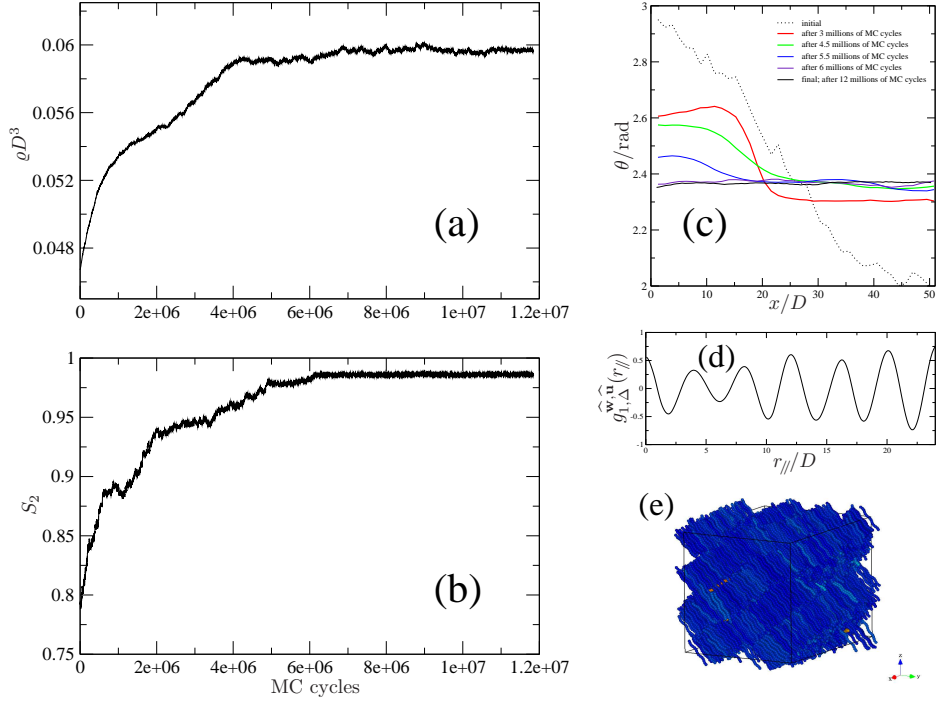} 
\caption{
Evolution of dimensionless number density (a) and 
nematic order parameter (b) as
a function of MC cycles in a system of hard helical particles with
$r=0.2D$ and $p=4D$ starting from 
a  cholesteric configuration at P$D^3/k_B$T=0.9 ; (c)
local nematic director angle $\theta$ as a function of $x$ at several
`instant' of MC `time' during the same calculation of panels (a) and (b);
(d) pair correlation function $g_{1,\Delta}^{{\widehat{\mathbf{w}}},{\widehat{\mathbf{u}}}}(r_{\parallelslam})$;
(e) image\cite{QMGA} of the system of hard helical particles 
with $r=0.2D$ and $p=4D$ in the screw-like smectic A phase 
as seen from the $yz$ plane normal; observe the layered structure 
in which these particles self-assemble and 
glimpse a very few of them lying in the inter-layer regions
transverse to the layer normal.
The hard helical particles are coloured according to the angle
that their $\widehat{\mathbf{u}}$ axes forms with
the eigenvector corresponding to the largest eigenvalue
of the global nematic order matrix.\cite{vieilliard}
}
\label{figevol}
\end{figure*}

The last representative case is given by helical particles with $r=0.2D$ and $p=4D$,
where both cholesteric and screw-like orderings are present (Fig. \ref{fig:summary}).
For these hard helical particle systems, 
a number of MC-NPT calculations were carried out at the following values
of P$D^3/k_B$T: 0.55; 0.6; 0.7; 0.8; 0.9. 
In previous numerical simulations 
employing 3D-PBC, 
the state points at P$D^3/k_B$T=0.55 and 0.6 
were assigned to a $\mathsf{N}$ phase.
Starting from 
a configuration obtained in the course of 
these previous numerical simulations and 
carefully putting up the two parallel hard walls,
the systems of hard helical particles with $r=0.2D$ and $p=4D$ too 
were relatively promptly seen to develop
a twist of their local nematic director, 
that is they were transforming to state points within a $\mathsf{N}^*_c$ phase,
as the relevant functions $g_2^{\widehat {\mathbf{u}}} (x)$ finally indicate (Fig. \ref{figP05506p}).
Both the sign and the order of magnitude of the corresponding $\cal{P}$ are found in agreement
with theoretical calculations,
though reaching a reasonable stationary value for the still negative $q$ required
lengthier runs than in the case of systems of hard helical particles with 
$r=0.2D$ and $p=9.92D$. 
With respect to the latter hard helical particle systems (Fig. \ref{figr02n1compost}), 
the functions $g_{1,\Delta}^{\widehat {\mathbf{w}},\widehat{\mathbf{u}}} \left( r_{\parallelslam} \right)$
none the less reveal an incipient screw-like ordering (Fig. \ref{figP05506p})
suggesting 
that it  could further develop at larger values of P$D^3/k_B$T.
\begin{figure*}[h!]
\centering
\psfrag{g1p rp }{$g_{1,\Delta}^{\widehat {\mathbf{w}},\widehat{\mathbf{u}}}\left( r_{\parallelslam} \right)$}
\psfrag{rp/D}{$r_{\parallelslam}/D$}
\psfrag{T/rad}{$\theta$/rad}
\psfrag{x/D}{$x/D$}
\psfrag{G1wrp}{$G_1^{\widehat{\mathbf{w}}}\left(x\right)$}
\includegraphics[scale=0.75]{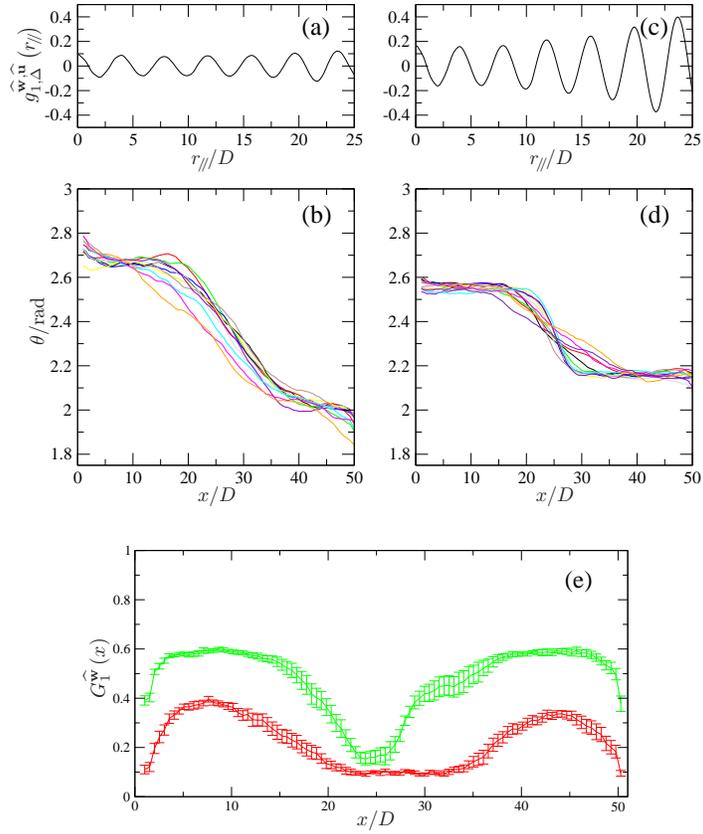}
\caption{Pair correlation functions $g_{1,\Delta}^{\widehat {\mathbf{w}},\widehat{\mathbf{u}}} \left( r_{\parallelslam} \right)$ and the angle $\theta$ that the local nematic director form with the $y$ axis
as a function of $x$ for a system of hard helical particles with
$r=0.2D$ and $p=4D$ as obtained in MC numerical simulations
starting from a nematic configuration inserted between two parallel hard walls
at  P$D^3/k_B$T=0.7 (a,b) and 0.8 (c,d). 
The various coloured thin full curves shown in (b,d) are each an 
average over 1000, one-thousand-MC-cycle separated, configurations. 
In (e) the function $G_1^{\widehat{\mathbf{w}}}\left(x\right)$ is also
shown at P$D^3/k_B$T=0.7 (red) and P$D^3/k_B$T=0.8 (green). 
}
\label{figP0708}
\end{figure*}
This is already the case at P$D^3/k_B$T=0.7 and 0.8
where the functions $g_{1,\Delta}^{\widehat {\mathbf{w}},\widehat{\mathbf{u}}} \left( r_{\parallelslam} \right)$
show an undamped cosinusoidal form 
[Fig. \ref{figP0708} (a), (c)],
in accord with previous numerical simulations
employing 3D-PBC that assigned
these two state points to a $\mathsf{N}_s^*$ phase.
Rather than starting from a screw-like nematic configuration
previously obtained at the same value of dimensionless pressure, 
these numerical simulations
were deliberately started from the same nematic configuration from which the present run
at  P$D^3/k_B$T=0.6 (Fig. \ref{figP05506p}) did start. 
If the behaviour of $g_{1,\Delta}^{\widehat {\mathbf{w}},\widehat{\mathbf{u}}} \left( r_{\parallelslam} \right)$
thus proves that screw-like ordering has spontaneously developed 
even once the system had been confined between two parallel hard walls,
the corresponding $\theta$ profiles [Fig. \ref{figP0708} (b), (d)] 
are not flat.
Rather, at both these values of dimensionless pressure, 
the system appears as having ended up blocked into
configurations characterised by two, overall flat, extremal regions
with a noticeable screw-like ordering,
separated and tied by an intermediate region of variable thickness
in which $\theta(x)$ accordingly drops and the
degree of screw-like ordering reduces [Fig. \ref{figP0708} (b), (d), (e)].
One natural interpretation of these results is that
each extremal region corresponds to a screw-like nematic film
wetting the respective hard wall. 
The further growing of these two films into one single domain could be
impeded for two alternative reasons. 
Either the intermediate region corresponds to an in-bulk
thermodynamically stable $\mathsf{N}_c^*$ phase and 
the two, overall flat, extremal regions are nothing more than
screw-like nematic films wetting the respective hard wall or 
the $\mathsf{N}_s^*$ phase is the in-bulk thermodynamically stable phase and 
the two extremal region misalignment takes very long to heal.
To help resolve this doubt, we resorted to the MD method 
to exploit its capability of naturally dealing with 
the system constituent particle collective motion.
Starting from a cholesteric configuration previously obtained at P$D^3/k_B$T=0.6, 
a lengthy MD-NPT calculation was carried out on 
a system of soft-repulsive helical particles confined between 
two parallel soft-repulsive walls at P$D^3/k_B$T=0.8.
The system loses its initial cholesteric character after 
$\sim$ 2 million of time steps and,
by forming also a central region with an overall flat
$\theta(x)$ profile, apparently 
progresses  towards a single-domain
untwisted screw-like nematic configuration: after
320 million time-steps,
this situation is so nearly reached that
one may confidently state that a neat $\mathsf{N}_s^*$ phase
is indeed forming starting from a cholesteric configuration (Fig. 
\ref{r02p4emdcompost}).
The latter untwisting process was eventually observed also 
at P$D^3/k_B$T=0.9 
in a system of hard helical particles confined
between two parallel hard walls.   
Starting from 
a cholesteric configuration
taken from the early stage of the run at P$D^3/k_B$T=0.7 and 
then letting the system evolve [Fig. \ref{figevol}(a) and (b)], 
a complete spontaneous untwisting of the local nematic director 
was finally detected [Fig. \ref{figevol}(c)]. This final phase is
screw-like (weakly) smectic A
as determined by calculating the smectic order parameter\cite{smeord} 
$\tau=0.25$, suitable pair correlation functions, 
as well as confirmed by direct visualisation [Fig \ref{figevol}(d,e)].
The same results were obtained starting from a moderately dense orthorhombic lattice
in which the hard helical particles  were perfectly aligned along the $z$ axis, 
the only differences being that reaching the equilibrium state was significantly faster 
and the degree of smectic ordering larger ($\tau=0.67$). These findings are consistent with those
previously obtained at the same dimensionless pressure by 
employing 3D-PBC.\cite{Kolli14b,Kolli16}
Starting from a screw-like smectic A configuration obtained in this latter run, 
additional MC-NPT calculations were carried out at P$D^3/k_B$T=0.8, 0.7 and 0.6. 
The flatness of the $\theta$ profile, the waviness of the  function 
$g_{1,\Delta}^{{\widehat{\mathbf{w}}},{\widehat{\mathbf{u}}}}(r_{\parallelslam})$ and 
the small value of $\tau=0.09$ are all indicative of a single-domain
$\mathsf{N}_s^*$ phase at P$D^3/k_B$T=0.8.
The same considerations would also hold at P$D^3/k_B$T=0.7
were it not for those fringes now visible at either hard walls,
a symptom that the $\mathsf{N}_c^*$ phase is close to forming.
\begin{figure}[h!]
\centering
\psfrag{x/D}{$x/D$}
\psfrag{rp /D}{$r_{\parallelslam}/D$}
\psfrag{g1p rp}{$g_{1,\Delta}^{{\widehat{\mathbf{w}}},{\widehat{\mathbf{u}}}}(r_{\parallelslam})$}
\psfrag{T/rad}{$\theta$/rad}
\includegraphics[scale=0.6]{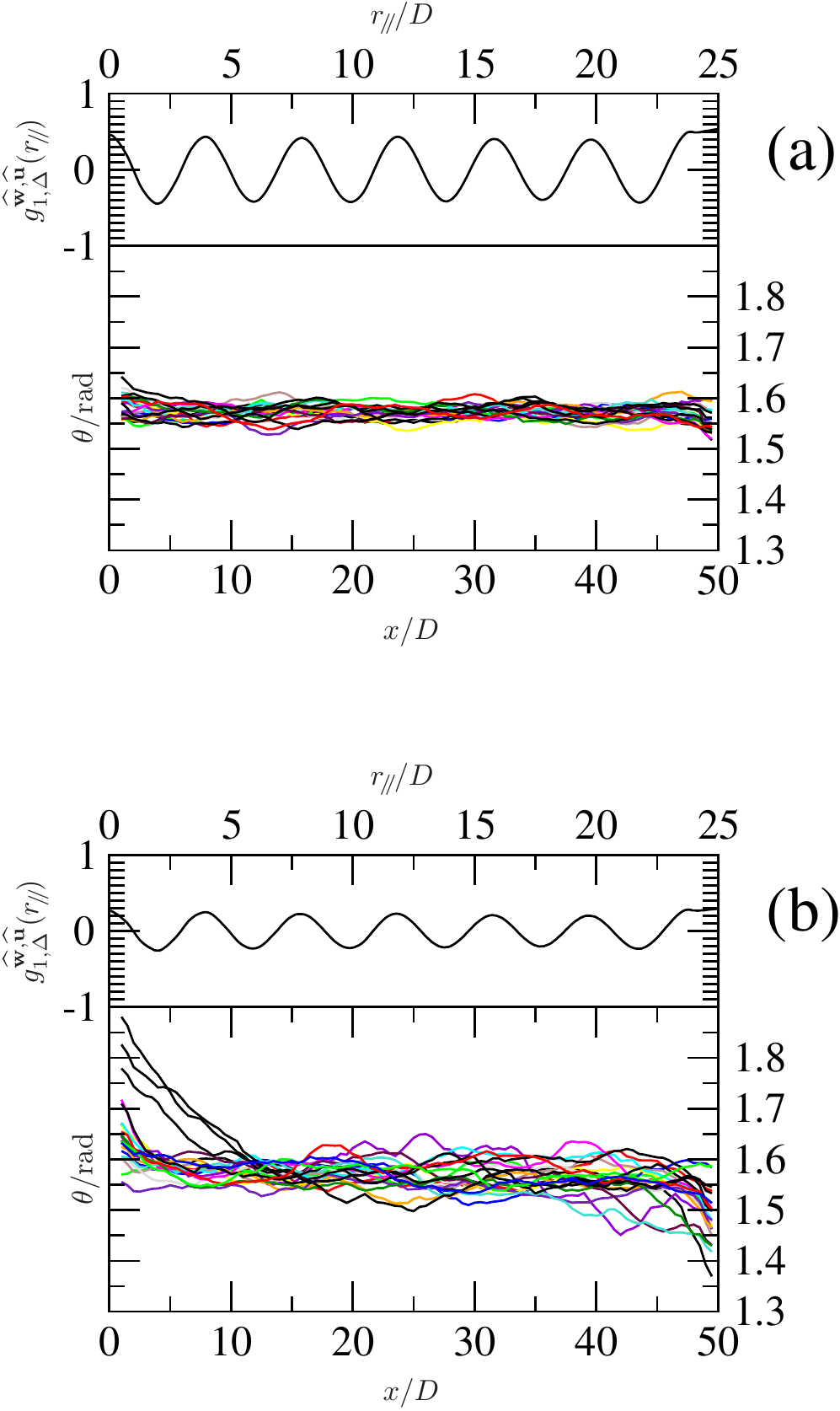}
\caption{Pair correlation functions 
$g_{1,\Delta}^{\widehat {\mathbf{w}},\widehat{\mathbf{u}}} \left( r_{\parallelslam} \right)$ (top panel) and 
the angle $\theta$ that the local nematic director forms with the $y$ axis
as a function of $x$ (bottom panel) for a system of hard helical particles with
$r=0.2D$ and $p=4D$ as obtained in MC numerical simulations
starting from a configuration obtained at P$D^3/k_B$T=0.9, 
in its turn obtained at the end of a run started from a moderately dense
orthorhombic lattice configuration inserted between two parallel hard walls
at P$D^3/k_B$T=0.8 (a) and 0.7 (b). 
The various coloured thin full curves shown in the bottom panel 
of (a,b) are each an 
average over 1000, one-thousand-MC-cycle separated, configurations.
}
\label{figP0807}
\end{figure}
In fact, it does at P$D^3/k_B$T=0.6,
consistent with the results also obtained by 
starting from a nematic configuration inserted between
the two parallel  hard walls (Fig. \ref{figP05506p}).

\section{Conclusions}
\label{sec:conclusions}
The aim of the present study was to address two important points 
originated from our past work on hard helical particle system. 
Firstly, the observability of the cholesteric phase.
(Elongated) hard helical particles are expected to form a cholesteric phase, 
but its pitch can be orders of magnitude longer than usual computational box dimensions and 
thus a proper cholesteric twist can remain unobservable using three-dimensional periodic boundary conditions.
Secondly, the stability of the screw-like nematic phase, a novel chiral nematic phase special to helical particles, 
that we observed in past work, against a proper cholesteric phase. 

By using isobaric-isothermal Monte Carlo and molecular dynamics methods, 
we have studied systems of helical particles 
confined between  two parallel repulsive walls 
so as to give a proper cholesteric phase the possibility to form.
Our findings confirm the existence of the two chiral nematic phases and are tersely summarised as:

 \begin{itemize}
  \item[I)] 
for weakly curly helical particles ($r=0.2D$ and $p=9.92D$), 
we observe a 
 cholesteric phase. 
 
  \item[II)] In the opposite limit of relatively curly helical particles  ($r=0.4D$ and $p=4D$), 
we find a  screw-like nematic phase  and no evidence of the cholesteric phase. 
 
  \item[III)] In the intermediate case  ($r=0.2D$ and $p=4D$), 
both the cholesteric and the screw-like nematic  phases are present at, respectively, lower and higher density.
  
  \end{itemize}
In principle,
both cholesteric and screw-like nematic phases can exist for a
system of helical particles but
the particle morphology controls the relative stability of these phases so that, for a given system, either may be absent.
For this reason,  phase diagrams and trends obtained without taking into account both cholesteric and screw-like nematic phases
should be regarded with caution.

The features of the cholesteric phase, when present, are in general agreement with current Onsager-like
theory results. 
For  $r=0.2D$, with both $p=4D$ and $p=9.92D$ a left-handed cholesteric phase is observed, as predicted by Straley \cite{straley}  
for right-handed helical particles with geometric parameters similar to those under study. 
Beside handedness, the Onsager-like theory calculations \cite{Frezza14,Belli14,Dussi15,Kolli16,tortora} provide 
reasonable estimates of the magnitude of the cholesteric pitch, of the order of 100-200$D$, 
which slightly decreases on going from $D=9.92$ to $D=4$.  
The decrease of the pitch  with increasing density found for the case $r=0.2D, p=9.92D$ is also in agreement with 
the theoretical calculations for $r=0.2D, p=8D$ in Ref. \onlinecite{Kolli16}.  
Indeed, a decrease of the cholesteric pitch, i.e. an increase of phase chirality, with 
increasing density is generally predicted for hard helical with  $p/D$ values larger than a few units. 
In Ref. \onlinecite{Dussi15,tortora} the opposite trend, i.e. an increase of the cholesteric pitch with increasing density, is reported for 
both   $r=0.4D, p=4D$ and $r=0.2D, p=4D$. 
However, such theoretical predictions refer to density ranges  
where the cholesteric phase does not seem to exist. 
Indeed, our numerical simulations do not show any cholesteric phase for the former helical particles, 
whereas for the latter the cholesteric range is so restricted that it is admittedly hard to identify 
a trend for the pitch as a function of density. 
In both these cases we have found the screw-nematic phase instead.\cite{Kolli14a,Kolli14b,Kolli16} 

The theoretical calculations for the cholesteric phase presented thus far\cite{Frezza14,Kolli16,Belli14,Dussi15,tortora} cannot account for screw-like nematic ordering.    
There is a need for a more general approach, as outlined in Ref. \onlinecite{libro}, to describe both chiral nematic phases.

Our present evidence suggests that the two chiral nematic phases are  mutually exclusive. 
However, we cannot fully exclude that cholesteric and truly long-ranged 
screw-like orderings might find a way to indeed coexist (e.g.
via the analogue of the twist-grain-boundary smectic phase\cite{TGB}).

\acknowledgments
G.C. thanks 
the Government of Spain for
the award of a Ram\'{on} y Cajal research fellowship and
acknowledges 
its financial support under 
grant FIS2013-47350-C5-1-R and grant MDM-2014-0377. 
This work was also supported by 
Italy MIUR via
PRIN-2010-2011 programme under grant 2010LKE4CC.
A.F. thanks C. Greco for the setup of the molecular dynamics numerical simulations.


\begin{thebibliography}{99}
\bibitem{comment}
The common tenet that the thermodynamic stability of a $\mathsf{N}_c^*$ phase is 
very close to that of a $\mathsf{N}$ phase providing $\cal{P}$ is sufficiently large \cite{dg}
motivated 
the usage of an Onsager-like  
(density functional) theory,\cite{onsager,pl}
within the implementation originally developed by Straley,\cite{straley}
to explore 
the characteristics of the $\mathsf{N}^*_c$ phase, 
in particular its pitch handedness and value, in systems of hard helical particles. 
\cite{Frezza14,Kolli16}
These results,
then reproduced also by others 
either using 
a different model \cite{wensink} or
the same model and a different implementation of the same theory,\cite{Belli14,Dussi15,tortora}
were useful in that they confirmed that  
phenomena such as cholesteric  
handedness inversion,
often explained by invoking a competition between 
different energetic and entropic terms that 
would promote opposite phase chirality, 
may arise also in pure systems of suitable hard chiral particles:
in such cases, they are solely entropy-driven.
One may none the less note  that cholesteric pitch handedness inversion, 
while possibly ``dramatic'' if viewed from the pitch side,
is a more ordinary phenomenon if viewed from the corresponding wavevector side:
it merely means that the  wavevector--versus--density function has a zero at a
certain density value and behaves linearly in its neighbourhood.
\end{thebibliography}
\end{document}